# Free Muons and Muonium - Some Achievements and Possibilities in Low Energy Muon Physics

Klaus P. Jungmann

*Kernfysisch Versneller Instituut,*
*Rijksuniversiteit Groningen, Zernikelaan 25,*
*NL 9747 AA Groningen, The Netherlands*

**Introduction**

The Standard Model (SM) is a theory framework, which allows an accurate description of all confirmed measurements in particle physics up to turn of the century. However, many observations are left without deeper explanations. Among those are the fundamental fermion mass spectrum, the origins of parity and CP violation, the fact of exactly three particle generations and many more. A variety of speculative models has been invented, in order to suggest physical interpretations of such yet not understood features contained in the SM.

Muons ($\mu^-$) and their antiparticles ($\mu^+$), the charged leptons in the second generation of fundamental fermions, have no internal structure down to dimensions of $10^{-18}$ m as shown in high energy lepton scattering experiments. They may therefore be regarded as point-like objects. The behaviour of muons can be described within standard theory with sufficient accuracy for all high precision experiments that have been carried out on them.

Muons are therefore important and central tools in a variety of research programs: The dominant $\mu^+$ decay into a positron ($e^+$), muon anti-neutrino and electron neutrino ($\mu^+ \rightarrow e^+ \bar{\nu}_\mu \nu_e$) yields the best value for the weak interaction Fermi coupling constant $G_F$. The insensitivity of muons to strong interactions makes muons important probes of nucleon properties in deep inelastic high-energy scattering. Muonic atom spectroscopy has given very reliable values for nuclear parameters, in particular nuclear charge radii. Searches for yet unobserved lepton number violating decays have yielded numerous bounds on crucial parameters in speculative models. High precision measurements of the electromagnetic interactions of free muons and such bound in the muonium atom (M = $\mu^+e^-$) – the hydrogen-like bound state of a positive muon and an electron ($e^-$) – have established stringent tests of standard theory, which includes in particular Quantum Electrodynamics (QED). The excellent agreements between measurements and this underlying theory has contributed significantly to today's view upon QED as the best available field theory. This solid confidence allows in return to extract most accurate values of fundamental constants such as the muon mass $m_\mu$, muon magnetic moment $\mu_\mu$ and magnetic anomaly $a_\mu$, and the electromagnetic fine structure constant $\alpha$.

*Muonium Production*

The best-known mechanism to produce muonium is $e^-$ capture after stopping $\mu^+$ in a suitable noble gas, where yields of 80(10)% can be achieved for krypton. Muons at accelerator facilities are born in weak pion decays and

In many muon experiments a limitation has been reached by now which is determined by the available particle fluxes at todays sources. For new accelerator facilities, such as the recently approved Japanese Hadron Project (JHP), a next generation of experiments has already been proposed. In addition to the exploitation of higher fluxes with established approaches novel techniques will be introduced to the field. This promises an expansion into new regions and can be expected to result in significant progress in exploring fundamental interactions and symmetries in physics (Table 1)[1].

In this article we will focus mainly on measurements of electromagnetic properties of the free muon and on spectroscopy of the muonium atom.

**Muonium**

The close confinement of the bound state in the muonium atom offers excellent opportunities to explore precisely fundamental electron-muon interactions. Since the effect of all known fundamental forces in this system are very well calculable within bound state QED, it renders both the possibility to extract precise constants as well as the possibility to search very sensitively for yet unknown interactions between these leptons. In contrast to natural atoms and ions as well as to artificial atomic systems, which contain hadrons, muonium has the advantage of being free of complications arising from the finite size and the internal structure of any of its constituents. Therefore, the system is particularly suited for searching new and yet unknown forces in nature.

In the muonium atom (Fig. 1) the most precise spectroscopy measurements can be performed on the ground state hyperfine structure splitting [2] and the 1s-2s energy interval [3]. Such experiments have been completed very recently after reaching limitations given by the quantities of available muons at the late Los Alamos Meson Physics Facility (LAMPF), in Los Alamos, USA, and at the Rutherford Appleton Laboratory (RAL) in Chilton, United Kingdom. These two transitions are experimentally favoured because they involve the 1s ground state in which the atoms can be produced with the highest yields. For accurate measurements the atoms need to be at low, ideally thermal velocities in the laboratory. The efficient conversion of an energetic $\mu^+$ beam into M is a key element in all experiments. parity violation in this process causes the muon beams to be polarized. The moderation processes involve dominantly the electric interaction and there is no muon depolarization. In strong axial magnetic fields (B >> 0.16 T) M is formed with a well defined ensemble average of the muon spin direction.

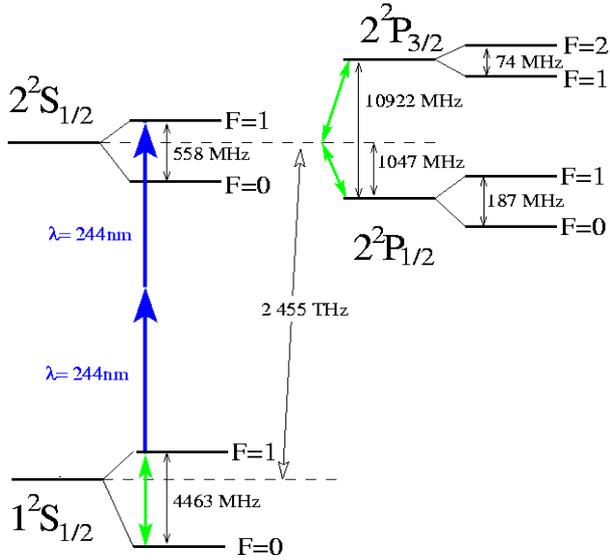

*Figure 1. Muonium energy levels for principal quantum numbesr n=1 and* n=2.

Muonium atoms at thermal velocities in vacuum can be obtained by stopping $\mu^+$ close to the surface of a target consisting of fine $SiO_2$ powder. The atoms are formed through $e^-$ capture and a fraction of a few percent of them diffuses through the target surface into the surrounding vacuum. Additional cooling with, e.g., laser techniques would not provide any significant advantages. Due to the $\tau_\mu = 2.2$ μs muon lifetime the natural line width of all transitions has a lower limit at $\Delta\nu_{nat} = (\pi \cdot \tau_\mu)^{-1} = 145$ kHz and cooling could not provide a much more advantageous line width. The described thermal M production technique has become an essential prerequisite for Doppler-free two-photon laser spectroscopy of the $1^2S_{1/2} - 2^2S_{1/2}$ interval $\Delta\nu_{1s2s}$ at KEK in Tsukuba, Japan, and with significantly higher precision at RAL. It was also the key to a sensitive search for a conversion of M into its anti-atom $\overline{M}$ at the Paul Scherrer Institut (PSI) in Villigen, Switzerland.

*Ground State Hyperfine Structure*
The most recent experiment at LAMPF had a Kr gas target inside of a microwave cavity at typically atmospheric density and in a homogeneous magnetic field of 1.7 T. Microwave transitions between the two energetically highest respectively two lowest Zeeman sublevels of the n=1 state at the frequencies $\nu_{12}$ and $\nu_{34}$ (Fig. 2) involve a muon spin flip. Due to parity violation in the weak interaction muon decay process the $e^+$ from $\mu^+$ decays are preferentially emitted in the $\mu^+$ spin direction. This allows a detection of the spin flips through a change in the spatial distribution of the decay $e^+$. As a consequence of the Breit-Rabi equation, which describes the behaviour of the M ground-state Zeeman levels in a magnetic field B, the sum of $\nu_{12}$ and $\nu_{34}$ equals at any strength of B the zero field splitting $\Delta\nu_{HFS}$. For sufficiently well known B the difference of these two frequencies yields the magnetic moment $\mu_\mu$.

The latest LAMPF experiment [2] has utilized the technique of "Old Muonium", which allowed to reduce the line width of the signals below half of the "natural" line width $\Delta\nu_{nat}$ (Fig. 3). For this purpose an essentially continuous muon beam was chopped by an electrostatic kicking device into 4 μs long pulses with 14 μs separation. Only decays of atoms which had been interacting coherently with the microwave field for periods longer than several muon lifetimes were detected.

The magnetic moment was measured to be $\mu_\mu$= 3.183 345 24(37) (120 ppb) which translates into a muon-electron mass ratio $m_\mu/m_e$ = 206.768 277(24) (120 ppb). The zero-field hyperfine splitting is determined to $\Delta\nu_{HFS}(exp)$ = 4 463 302 765(53) Hz (12 ppb) which agrees well with the theoretical prediction of $\Delta\nu_{HFS}(theo)$ = 4 463 302 563(520)(34)(<100) Hz (120 ppb). Here, the first quoted uncertainty is due to the accuracy to which $m_\mu/m_e$ is known, the second error is from the knowledge of $\alpha$ as extracted from Penning trap measurements of the electron magnetic anomaly, and the third uncertainty corresponds to estimates of uncalculated higher order terms. Among the non QED contributions is the strong interaction through vacuum polarization loops with hadrons which adds 250 Hz and a parity conserving axial vector–axial vector weak interaction which is −65 Hz.

For the muonium hyperfine structure the comparison between theory and experiment is possible with almost two orders of magnitude higher precision than for natural hydrogen because of the not sufficiently known proton charge and magnetism distributions. For hydrogen the achieved some six orders of magnitude higher experimental precision (in hydrogen maser experiments) can therefore unfortunately not be exploited for a better understanding of fundamental interactions.

Among the possible exotic interactions, which could contribute to $\Delta\nu_{HFS}$, is muonium-antimuonium conversion [4] (see below). Here, an upper limit of 9 Hz could be set from an independent experiment described below. Recently, generic extensions of the SM, in which both Lorentz invariance and CPT invariance are not assumed, have attracted widespread attention in physics. Diurnal variations of the ratio $(\nu_{12} - \nu_{34})/(\nu_{12} + \nu_{34})$ are predicted. An upper limit could be set from a reanalysis of the LAMPF data at $2\times10^{-23}$ GeV for the Lorentz and the particle mass. In this framework $\Delta\nu_{HFS}$ provides a CPT violating parameter. In a specific model by Kostelcky and co-workers a dimensionless figure of merit for CPT tests is sought by normalizing this parameter to *[2]*.significantly better test of CPT invariance than electron g-2 and the neutral Kaon oscillations [5].

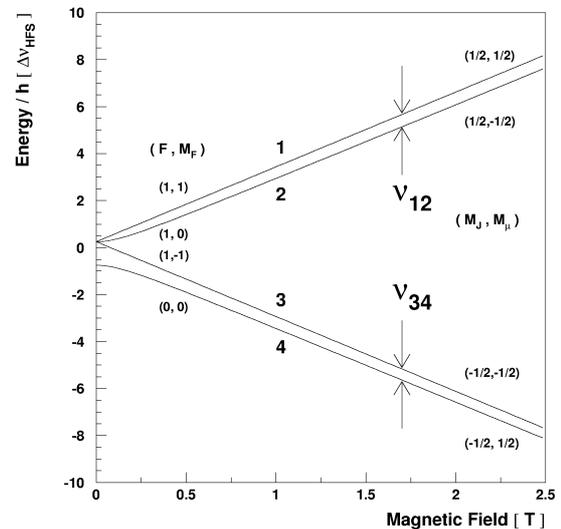

*Figure 2. Muonium ground state hyperfine structure Zeeman splitting.*

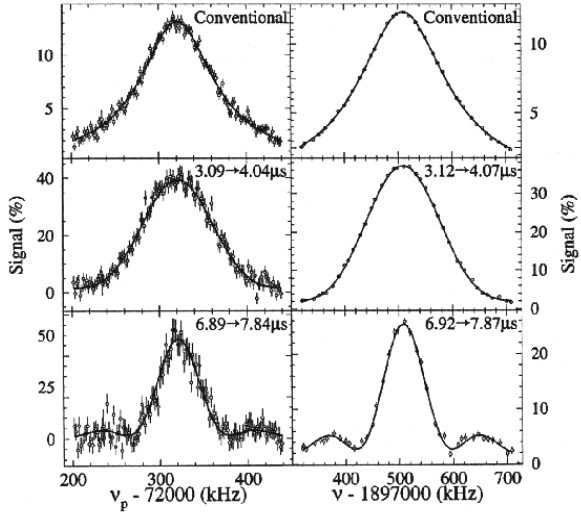

*Figure 3. Samples of conventional and "old muonium" resonances at frequency $v_{12}$. The narrow "old" lines exhibit a larger signal amplitude. The signals were obtained with magnetic field sweep (left column, magnetic field in units of proton NMR frequencies) and by microwave frequency scans (right column)*

The hyperfine splitting is proportional to $\alpha^2 \cdot R_\infty$ with the very precisely known Rydberg constant $R_\infty$. Comparing experiment and theory yields $\alpha^{-1}$ = 137.035 996 3(80) (58ppb). If $R_\infty$ is decomposed into even more fundamental constants, one finds $\Delta v_{HFS} \propto \alpha^4 \cdot m_e/h$. With $h/m_e$ as determined in measurements of the neutron de Broglie wavelength we have $\alpha^{-1}$ = 137.036 004 7(48) (35 ppb). In the near future a small improvement in this figure can be expected from ongoing determinations of $h/m_e$ in measurements of the photon recoil in Cs. A better determination of the muon mass, e.g. will result in a further improvement and may contribute to resolving the situation of various poorly agreeing determinations of the fine structure constant, which is important in many different fields of physics.

It should be mentioned that the present agreement between $\alpha$ as determined from M hyperfine structure and from the electron magnetic anomaly is generally considered the best test of internal consistency of QED, as one case involves bound state QED and the other one QED of free particles.

The results from the LAMPF experiment are mainly statistics limited and improve the knowledge of both $\Delta v_{HFS}$ and $\mu_\mu$ by a factor of three over previous measurements. This gain could be significantly surpassed at a future high flux muon source.

### *The 1s-2s interval in muonium*

In muonium the 1s-2s energy difference is essentially given by the relevant quantum numbers, $R_\infty$ and a reduced mass correction. Therefore, this transition may be regarded as ideal for a determination of the muon-electron mass ratio. QED corrections are well known for the needs of presently possible precision experiments and do not play an important role here. Doppler-free excitation of the 1s-2s transition has been achieved in pioneering experiments at KEK and at RAL. In all these experiments two counter-propagating pulsed laser beams at 244 nm wavelength were employed to excite the n=2 state. The successful transitions were then detected by photo-ionization with a third photon from the same laser field. The released $\mu^+$ was then registered on a microchannel plate detector.

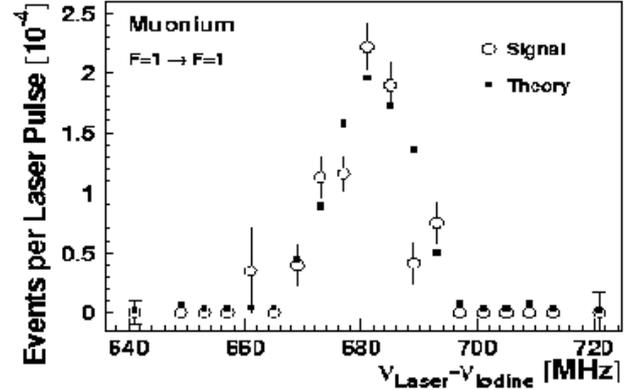

*Figure 4. Muonium 1s-2s signal. The frequency scale corresponds to the offset of the laser system base frequency from a molecular iodine reference line. The open circles are the observed signal, the solid squares represent the theoretical expectation based on pulse-by-pulse measured laser beam para*meters *(phase and intensity) and a line-shape model [3,6].*

The accuracy of the early measurements was limited by the ac-Stark effect and rapid phase fluctuations (frequency chirps), which were inherent properties of the necessary pulsed high power laser systems. The key feature for the latest high accuracy measurement at RAL was a shot by shot measurement of the spatial laser intensity profile as well as the time dependences of the laser light intensity and phase. This together with a newly developed theory of resonant photo-ionization [6] allowed a shot-by-shot prediction of the transition probability as a basis for the theoretical line shape (Fig. 4).

The latest RAL experiment [3] yields $\Delta v_{1s2s}(exp)$= 2 455 528 941.0(9.8) MHz in good agreement with a theoretical value $\Delta v_{1s2s}(theo)$=2 455 528 935.4(1.4) MHz. The muon-electron mass ratio is found to be $m_{\mu^+}/m_{e^-}$ = 206.768 38(17). Alternatively, with $m_{\mu^+}/m_{e^-}$ as extracted from the M hyperfine structure, a comparison of experimental and theoretical values can be interpreted in terms of a $\mu^+$ - $e^-$ charge ratio, which results as $q_{\mu^+}/q_{e^-}$ +1= −1.1(2.1) × $10^{-9}$. This is the best verification of charge equality in the first two generations of particles. The existence of one single universal quantised unit of charge is solely an experimental fact and no underlying symmetry could yet be revealed. The interest in such a viewpoint arises because gauge invariance assures charge quantization only within one generation of particles.

Major progress in the laser spectroscopy of M can be expected from a continous wave laser experiment, where frequency measurement accuracy does not present any problem because light phase fluctuations are absent. For this an intense source of muons will be indispensable.

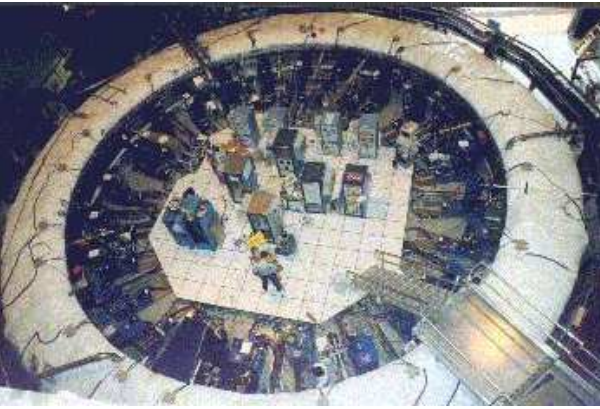

*Figure 6. The muon g-2 storage ring experiment at BNL*

*Muon Magnetic Anomaly*

The spectroscopy experiments on muonium and a measurement of the muon magnetic anomaly $a_\mu = (g_\mu-2)/2$ are closely related through $\mu_\mu = g_\mu\, e\hbar/(2m_\mu c)$. This arises from the precise values of fundamental constants and the high accuracy tests of the validity and reliability of QED for leptons, which both form an indispensable basic input for the analysis of the measured data and the calculations of a theoretical value. The measurements in muonium spectroscopy and of $a_\mu$ together put a stringent test on the internal consistency of theory and the values of the involved constants $a_\mu$, $m_\mu$, $\mu_\mu$ and $\alpha$.

The muon magnetic anomaly has been measured in three past experiments at CERN to 7 ppm. The anomaly arises from interactions with virtual particles created by the muons own radiation field. It is dominated, like in case of the electron, mostly by virtual electron, positron and photon fields. However, the effects of heavier particles are enhanced in comparison to the electron case by the square of the mass ratio $m_\mu/m_e \approx 4\times 10^4$. Whereas for the electron such contributions altogether amount to about the present experimental uncertainty at 4 ppb, they have been experimentally demonstrated for muons already clearly in the last CERN experiment. The influence of the strong interaction can be determined in its dominating first order vacuum polarization part from a dispersion relation with the input from experimental data on $e^+$-$e^-$ annihilation into hadrons and hadronic $\tau$-decays. It amounts to 58 ppm. Part of the hadronic contributions is hadronic light-by-light scattering. This can only be determined from calculations and is the subject of ongoing highly actual research [7]. Even the sign of the effect has frequently changed in calculations within the past decade. The weak interaction adds 1.3 ppm. Accounting for all known effects, present standard theory yields $a_\mu$ to order 1 ppm. Possible influence from physics beyond the SM may be as large as a few ppm. Such could arise, for example, from supersymmetry, compositeness of fundamental fermions and bosons, CPT violation and from many others.

There is a twofold high value for a precision measurement of $a_\mu$. Firstly, a discrepancy with finally agreed and confirmed standard theory calculations would give hints to yet undiscovered interactions and particles and it would stimulate more direct searches. Secondly, a good agreement at a high level of accuracy would set stringent limits on parameters in a large number of speculative models.

A new determination of $a_\mu$ is presently carried out in a superferric magnetic storage Ring at the Brookhaven National Laboratory (BNL) in Upton, USA (Fig. 5) [8]. The difference between the spin precession and the cyclotron frequencies of the stored muons is determined. The detailed analysis of all data obtained until the year 2000 with together some 5 billion $\mu^+$ has gien a combined experimental value of $a_\mu(\exp)= 11659\ 203(8)\times 10^{-10}$ (0.7 ppm). In 2001 data were collected from some 4 billion $\mu^-$ decays. The BNL program committee has approved further running with $\mu^-$ from which one expects additional 6 billion events. This will allow to achieve comparable accuracy for both signs of muon charge as a sensitive test of CPT invariance.

Most recently two precise but different values have been calculated on the basis of standard theory by a single theoretical group [7]. These results on whether $\tau$ decay data were used or not for determining the hadronic contributions. They differ from the measured value by 1.6 times respectively 3.0 times the combined experimental and theoretical uncertainties. An earlier reported larger difference [8] had led to a careful review of all SM contributions. In this process a calculational error was found in hadronic light-by-light scattering, showing the sensitivity to the precision of calculations and uncertainty assignments for theoretical values. This is work in progress and the future will show whether a the final result will be a hint to new physics beyond the SM.

It should be noted that there is a severe limitation to the interpretation of a perhaps future muon $g_\mu$-2 measurement, in connection with extracting or limiting parameters of speculative models, which arises from the hadronic vacuum polarization and owes to the fact that measurements of $e^+$-$e^-$ annihilation into hadrons and hadronic $\tau$ decays have reached a level of precision not too far from the limits which arise from the statistics achievable at present facilities.. Hadronic light-by-light scattering, which only can be taken from calculations, sets a principal limit as long as the associated conceptual problems remain unsolved.

In order to find new physics in precision measurements it may therefore be advantageous to use systems, where the standard theory predictions are simpler. Searches for a permanent electric dipole moment (edm) of any fundamental particle are here good examples. Electric dipole moments are forbidden by P, T and CP invariance and the SM predictions are several orders of magnitude below present search limits. Furthermore, such research gains additional motivation, because the identification of new sources of CP violation could be a crucial ingredient for explaining the dominance of matter over antimatter in the Universe. Driven by these arguments a novel idea has been brought forward [9]. It is to search for a muon edm by exploiting the motional electric field a highly elativistic muon experiences in a magnetic field. This motional field can be orders of magnitude stronger than technically achievable fields. A six orders of magnitude improvement over the present limit is aimed for by a collaboration at BNL. At this level a muon edm experiment will be competitive with electron or neutron experiments and has the further benefit of probing a new particle generation.

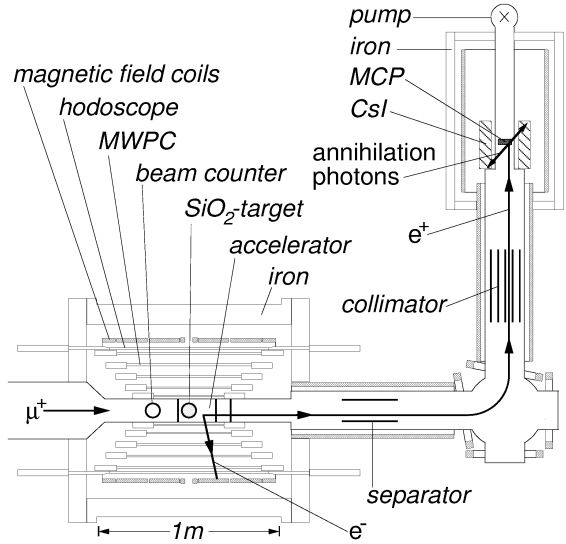

*Figure 6. Muonium–Antimuonium Conversion Spectrometer at PSI*

*Muonium to Antimuonium Conversion*

In addition to the indirect searches for signatures of new physics in the muon magnetic anomaly and in electromagnetic interactions within the muonium atom the bound state offers also the possibility to search more directly for predictions of speculative models. The process of muonium to antimuonium-conversion violates additive lepton family number conservation. It would be an analogy in the lepton sector to $K^0\overline{K}^0$ oscillations.. $M\overline{M}$-conversion appears naturally in many theories beyond the SM. The interaction could be mediated, e.g., by a doubly charged Higgs boson $\Delta^{++}$, Majorana neutrinos, a neutral scalar, a supersymmetric τ-sneutrino $\tilde{\nu}_\tau$, or a doubly charged bileptonic gauge boson $X^{\pm\pm}$.

At PSI an experiment was designed to exploit a powerful new signature, which requires the coincident identification of both particles forming the anti-atom in its decay [4]. Thermal muonium atoms in vacuum from a $SiO_2$ powder target, are observed for $\overline{M}$ decays. Energetic electrons from the decay of the $\mu^-$ in the $\overline{M}$ atom can be identified in a magnetic spectrometer (Fig. 6). The positron in the atomic shell of $\overline{M}$ is left behind after the decay with 13.5 eV average kinetic energy. It can be post-accelerated and guided in a magnetic transport system onto a position sensitive microchannel plate detector (MCP). Annihilation radiation can be observed in a segmented pure CsI calorimeter around it. The decay vertex can be reconstructed.

The measurements were performed during a period of 6 months in total over 4 years during which $5.7 \times 10^{10}$ M atoms were in the interaction region. One event fell within a 99% confidence interval of all relevant distributions. The expected background due to accidental coincidences is 1.7(2) events. Depending on the interaction details one has to account for a suppression of the conversion in the 0.1 T magnetic field. This amounts maximally to a factor of about 3 for V±A type interactions. Thus, the upper limit on the conversion probability is $8.2 \times 10^{-11}$ (90% C.L.). The coupling constant is bound to below $3.0 \times 10^{-3}$ $G_F$.

This new result, which exceeds limits from previous experiments by a factor of 2500 and one from an early stage of the experiment by 35, has some impact on speculative models. For example: A certain $Z_8$ model is ruled out which has more than 4 generations of particles and where masses could be generated radiatively with heavy lepton seeding. A new lower limit of $m_{X^{\pm\pm}} \leq 2.6$ TeV/$c^2 * g_{3l}$ (95% C.L.) on the masses of flavour diagonal bileptonic gauge bosons in GUT models is extracted, which lies well beyond the value derived from direct searches, measurements of the muon magnetic anomaly or high energy Bhabha scattering. Here, $g_{3l}$ is of order unity and depends on the details of the underlying symmetry. For 331 models the experimental result can be translated into $m_{X^{\pm\pm}} \leq 850$ GeV/$c^2 * g_{3l}$ which excludes some of their minimal Higgs versions, where an upper bound of 600 GeV/$c^2$ has been extracted from an analysis of electro-weak parameters. The 331 models need now to refer to a less attractive and more complicated extensions. In the framework of R-parity violating supersymmetry the bound on the relevant coupling parameters could be lowered by a factor of 15 to $\lambda_{132} \cdot \lambda_{231} * 3 \times 10^{-4}$ for assumed superpartner masses of 100 GeV/$c^2$.

A future $M\overline{M}$-experiment could particularly take advantage of high intensity pulsed beams. In contrast to other lepton number violating muon decays, the conversion through its nature as particle-antiparticle oscillation has a time evolution in which the probability for finding a system formed as M decaying as $\overline{M}$ increases quadratically in time. This gives the signal an advantage, which grows in time over exponentially decaying background. E.g., with a twofold coincidence as part of a signature after $\Delta T = 2\tau_\mu$ beam related accidental background has dropped by almost two orders of magnitude, whereas a $M\overline{M}$-signal would not have suffered significantly at all.

**Future possibilities**

All precision muonium experiments are now limited by statistics. For this reason significant improvements can be expected from more efficient M atom formation. There, are some encouraging developments at RIKEN-RAL where hot metal targets are used [10]. However, by far the most promising approaches are muon sources with higher intensities. Such may become available, in principle, at any high power proton facility with particle energies above the pion production threshold. The dominating figure of merit is the beam power on the production target. At JAERI in Japan the construction of the Japanese Hadron Project (JHP) has been started which has a 1 MW proton beam. Further, novel muon beam line concepts use compared to present facilities much larger particle collection solid angles at the production target and aim for significant phase space cooling of the beam. Examples are the DIOMEGA and PRISM projects [10].

| Type of Experiment | Physics Issue | Possible Experiments | Present Accuracy | Possible Future Accuracy |
|---|---|---|---|---|
| "Classical" Rare and Forbidden decays | lepton number violation; New Physics searches | $\mu^+ e^- \to \mu^- e^+$ <br> $\mu \to e \gamma$ <br> $\mu \to eee$ <br> $\mu^- N \to e^- N$ | $8.1 \times 10^{-11}$ <br> $1.2 \times 10^{-11}$ <br> $1.0 \times 10^{-12}$ <br> $6.1 \times 10^{-13}$ | $< 10^{-13}$ (novel concept) <br> $< 10^{-15}$ (novel techniques) <br> $< 10^{-16}$ (novel techniques) <br> $< 10^{-18}$ (novel method) |
| Muonium Spectroscopy | fundamental constants $m_\mu$, $\mu_\mu$, $\alpha$, $q_\mu$; weak interactions, muon charge | $\Delta\nu_{HFS}$ <br> $\Delta\nu_{1s2s}$ | $12 \times 10^{-9}$ <br> $1 \times 10^{-9}$ | $< 5 \times 10^{-9}$ (exploit. statistics) <br> $< 10^{-11}$ (novel concept) |
| Muon Moments | Standard Model tests; New Physics searches; T, CP, CPT tests | $g_\mu - 2$ <br> $\mu$ electric dipole moment | $1.3 \times 10^{-6}$ <br> $3.4 \times 10^{-19}$ ecm | $< 10^{-7}$ (exploit. statistics) <br> $< 5 \times 10^{-26}$ ecm (novel concept) |
| Muonic Atoms | nuclear parameters; nuclear charge radii; weak interactions | $\mu^-$ atoms <br> radioactive $\mu^-$ atoms | depends on system not yet performed | depends on system (novel ideas involving particle traps) |

**Table 1.** Some muon physics experiments where significantly enhanced accuracy can be expected at intensive muon sources. The prospected future accuracies have been estimated for a (pulsed) 4 MW proton accelerator facility [1]. The gain in precision arises not only from improved statistics. In many cases novel concepts and techniques can be applied.

In Europe a spallation source and a neutrino factory are being discussed. Also, the planned new GSI machine could provide such beams, if rapid cycling would be foreseen. In a similar way the Brookhaven AGS could be upgraded. The success of such high power facilities will crucially depend on the capability of the possible targets to withstand high beam powers. Therefore, strong research activities should be focussed on this aspect soon.

In addition to more precise measurements in muonium a rich variety of experiments could be served at such expanded facilities. In Table 1 some possibilities are given, which include spectroscopy of artificial atoms and ions like muonic hydrogen and muonic helium where important parameters describing the hadronic particles within these systems can be determined. At such new facilities in particular several novel experimental techniques will become feasible and can be expected to provide most sensitive tests of fundamental interactions in an area with a high potential to discover new physics.

**Conclusions**

Professor I.I. Rabi's question "Who ordered that ?", after he had learned about the muon being a heavy lepton, has not been answered yet. The nature of the muon - the reason for its existence - still remains an intriguing mystery to be solved. On the way to find an answer, theorists and experimentalists have contributed through their complementary work in fundamental muon physics to an improved understanding of basic particle interactions and fundamental symmetries in physics. Particularly muonium spectroscopy has verified the nature of the muon as a point-like heavy lepton, which differs only in mass, related parameters from the electron (and the tau). In addition, these measurements have provided accurate values of fundamental constants. With new high flux machines a fruitful future must be expected.